\documentclass[pdflatex,sn-mathphys-num]{sn-jnl}

\usepackage{graphicx}%
\usepackage{multirow}%
\usepackage{amsmath,amssymb,amsfonts}%
\usepackage{amsthm}%
\usepackage{mathrsfs}%
\usepackage[title]{appendix}%
\usepackage{xcolor}%
\usepackage{textcomp}%
\usepackage{manyfoot}%
\usepackage{booktabs}%
\usepackage{algorithm}%
\usepackage{algorithmicx}%
\usepackage{algpseudocode}%
\usepackage{listings}%
\usepackage{multirow}
\usepackage{xcolor}
\usepackage{cleveref}

\definecolor{RoyalBlue}{RGB}{65,105,225}
\definecolor{FireBrick}{RGB}{178,34,34}
\definecolor{ForestGreen}{RGB}{34,139,34}

\theoremstyle{thmstyleone}%
\theoremstyle{thmstyletwo}%
\usepackage{float}
\usepackage{cleveref}
\theoremstyle{thmstylethree}%
\usepackage{geometry}
\begin{document}


\title{A geometric physics-informed machine learning inference for the neutron star maximum mass and the inverse problem}


\author[]{\fnm{Rounak} \sur{Mukherjee}}\email{rounak21@iiserb.ac.in}

\author[]{\fnm{Ritam} \sur{Mallick}}\email{mallick@iiserb.ac.in}

\affil[]{\orgdiv{Department of Physics}, \orgname{Indian Institute of Science Education and Research Bhopal}, \orgaddress{\city{Bhopal}, \postcode{462066}, \country{India}}}

\abstract{The existence of a distinct mass boundary between the heaviest neutron stars and the lightest black holes remains in question. It is an artefact of our ignorance of the properties of matter at supra-nuclear densities, which exist in the cores of neutron stars. The study addresses these problems with a physics-informed machine learning approach, guided by astrophysical observations. The Transformer model is trained on an agnostically generated ensemble of equations of state. Two geometric parameters are defined on the mass-radius sequence of a neutron star--the front bending and the back bending. The transformer provides a two-step solution: first, the model predicts the maximum mass and radius using the bending parameters. Second, it predicts the square of the sound speed profile, completing the inverse mapping. The prediction is that massive neutron stars form when the sound speed peaks at low density, leading to strong back-bending and an early phase transition to quark matter. Massive stars favour a stiff equation of state at low density, and the density of matter at the star's core is sufficiently small. The maximum mass for a neutron star predicted by the astrophysical constrained transformer model is $2.477$ solar masses, and a minimum radius of about $11.498$ km for a neutron star of $1.4$ solar masses.}


\maketitle

\section{Introduction}\label{sec1}

White dwarfs (WDs), neutron stars (NSs), and Black holes (BHs) constitute the majority of the astrophysical compact objects one encounters. Depending on the mass of the progenitor main-sequence star, they ultimately collapse to one of the compact objects after running out of their nuclear fuel \cite{Shapiro:1983du, Haensel:2007yy}. These objects are extremely complex yet provide a unique environment that does not exist anywhere else in the universe. Among them, NSs provide an interesting laboratory for analysing both matter and gravity under extreme conditions \cite{Shuryak:1980tp, 1996Glendenning}. As the gravity is extremely high at the cores of NSs, the matter reaches extremely high densities, which cannot be replicated in terrestrial laboratory settings \cite{Lattimer:2004pg, Annila:2023axs, Burgio:2021vgk, Baym_2018}. Therefore, NSs serve as laboratories to test theories of gravity, as well as particle and nuclear physics.

One of the most intriguing features of compact stars, which remains shrouded in mystery, is mass segregation. Although a well-defined theoretical limit on the maximum mass of WDs exists \cite{Chandrasekhar1931ApJ}, recent observations have found few compact objects lying in the intermediate mass-radius region between NSs and WDs, the HESS and XTE \cite{Baglio:2013uoa, Doroshenko2022}. On the other end, the maximum mass of NS is poorly defined, and a significant gap exists between NSs and BHs, the so-called \textit{mass gap} \cite{Rhoades, lindbolm, Lattimer_Prakash}. However, recent gravitational wave (GW) observations have identified a few objects that fall within this mass gap, like GW190814, GW200105, GW230529 \cite{2017GW170817, Abbott_2020_GW190425, Abbott_2020_GW190814, 2021ApJ915, Abac_2024, Kacanja_2025}. Depending on the model, these objects are predicted to be either extremely massive NSs or extremely small BHs \cite{Abbott_2020_GW190425, Abbott_2020_GW190814, Abac_2024}. However, the exact nature of these objects still remains unclear.

The quest for the maximum mass of NS or the minimum mass of BHs is not new. It started in the 1970s \cite{Rhoades}, where a rough estimate of the maximum mass of NS was predicted employing the maximum bound on the sound speed (causality). A similar line of enquiry later projected the maximum redshift and maximum compactness of an NS \cite{lindbolm, Lattimer_Prakash}. Most of the NSs observed in the late twentieth century (as pulsars) were around $1.4$ solar mass; whereas the smallest BHs observed were around 5 solar masses \cite{Thompson2019}, resulting in the emergence of the notion of the \textit{mass gap}\cite{Mallick_2019, Mallick:2020zpd, Ecker:2022dlg, Musolino:2023edi, Saha:2024vst, Rezzolla:2025pft}.

The scenario changed drastically in the twenty-first century with the phenomenal advancement of observational astrophysics. Early equation of state (EoS) models at high density were significantly soft to match the canonical mass of NSs ($1.4$ solar mass), but were soon discarded with the discovery of several massive pulsars \cite{_zel_2010, Demorest_2010, Antoniadis_2013, Arzoumanian_2018, Cromartie_2019}. In the meantime, two important observatories changed the overall understanding of NSs and high-density matter: NICER and LIGO-VIRGO-KAGRA. The detection of the binary neutron star merger GW170817 via gravitational waves \cite{2017GW170817} provided a low upper cut-off on the tidal deformability of the NSs \cite{Bauswein_2017, Raithel:2018ncd, De_2018, Rezzolla_2018, Annala_2018, 2018PhRvL.121p1101A}. The NSs that merged had masses between 1.1 and 1.4 solar masses, and thus the bound on tidal deformability pointed towards a softer EoS for intermediate-mass NSs. The mass-radius measurement of PSR J0437--4715 \cite{choudhury_nicer_2024} and PSR J0614--3329 \cite{Mauviard2025} indicated a softer EoS for intermediate mass NSs, and on the other hand the measurements of J0030+0451 \cite{Riley_2019,Miller_2019,Raaijmakers_2019} and J0740+6620 \cite{Riley_2021,Miller_2021,Raaijmakers_2021,Fonseca_2021} suggested a stiffer EoS for higher masses. Therefore, the most probable EoS for NS is surely non-monotonous, but a definitive answer still eludes \cite{2025ApJ...988..258V}.

The maximum mass of NSs, therefore, depends on the interplay between the details of the EoS \cite{Komoltsev:2023zor} at various densities found in them and the gravity compressing the stars \cite{Saha:2025don}. Therefore, to apprehend the totality of the EoS found in NS interiors, measurement of a single NS is not enough, and one needs a wider range of observation of their mass-radius sequence (M-R) \cite{Steiner:2010fz, Bednarek:2011gd, Steiner:2012xt, Sieniawska:2018pev, Tews_2018, Jiang:2022tps, Ecker:2024uqv, Saha:2024swd, Tewari:2024qit, Ferreira:2025dat, Li:2025uaw, Verma:2025vkk,  Saha:2025don}. This requires precise measurements of several pulsars across different mass ranges, which is the primary objective of NICER x-ray missions, but may take considerable time to realise in practice. Therefore, given the present observation, one can recognise some characteristic features from the M-R and solve the so-called \textit{inverse problem} for NSs, where one tries to infer the EoS from the mass-radius sequence \cite{1992ApJ...398..569L, Fujimoto_2020, Morawski:2020izm, Soma:2022vbb, Soma:2023rmq, Carvalho:2025qie, Patra:2026yuv}. This is a non-trivial problem having large degeneracies and poses severe challenges. However, with the advancement of machine learning, we address both \textit{inverse problem} and \textit{mass gap} by identifying a novel geometric property of the mass-radius sequence: \textit{front-bending (FB)} and \textit{back-bending (BB)}. These two geometric parameters provide an efficient way to draw inferences from the M-R space, connecting the NS observations to the EoS space and narrowing down the large possibilities of the M-R space.


These two parameters intricately relate the M-R sequence to the maximum mass and the square of the sound speed ($c_s^2$) \cite{Altiparmak:2022bke, Chatterjee:2023ecc, Verma:2026npn}. The machine learning algorithm is trained on an ensemble of agnostically generated EoS to predict the maximum mass. We further use the maximum mass and radius as input to solve the \textit{inverse problem} and obtain the governing EoS. Unlike conventional approaches to unravel the dense matter EoS, where an EoS generates M–R sequences and the inverse inference remains highly degenerate due to the infinite family of M–R curves consistent with observations—our framework performs a direct inverse mapping governed by the intrinsic geometry of the M–R sequence, that is bound within the finite value of the proposed parameters, enabling us to tackle this degeneracy problem. Additionally, these two parameters, along with the characteristic points, can represent the entire M-R sequence quite efficiently. 

The paper is organised as follows: in Section II, we provide a detailed description of our formalism for EoS generation, including the formulation of back-bending and front-bending, as well as the transformer model. In Section III, we present our results on the maximum mass of NSs given the observational bounds. Finally, in Section IV, we summarise and draw important conclusions from our results.

\section{Formalism}

The \textit{inverse method} in principle should scan the M-R sequence and reproduce the actual EoS. Presently, there exists an infinite number of possible M-R sequences that are able to satisfy all the NICER and GW constraints. Therefore, it is presently impossible to obtain a unique M-R sequence and thereby a unique EoS prediction by solving the \textit{inverse problem}. One can therefore have a number of possible M-R sequences and an ensemble of EoS that most likely satisfy all the present constraints. 

To address the problem, the machine trains on the regular problem, where one solves the Tolman-Oppenheimer-Volkoff (TOV) equation for a given EoS \cite{Oppenheimer1939, Tolman1939}. As the EoS is not known a priori, one generates an ensemble of agnostic EoS using a speed-of-sound parametrisation, maintaining thermodynamic consistency \cite{2022ApJ...939L..34A, 2023NatCo..14.8451A, 2025ApJ...988..258V}. From very low density till $0.5$ times the nuclear saturation density ($n_0$), the EoS uses a BPS crust \cite{1971NuPhA.175..225B}. Chiral effective field theory (CET) construction is further applied to achieve a narrow band up to $1.1 n_0$ \cite{2013ApJ...773...11H}. Beyond the given density, the sound speed interpolation method is used to construct the EoS. The sound speed is parametrised as a function of baryon chemical potential for five randomised segments bridging the CET chemical potential with the perturbative QCD potential \cite{2014ApJ...789..127K}. Linear interpolation is further implemented to generate a continuous curve. We generate continuous curves of two categories, monotonic and non-monotonic \cite{2025ApJ...988..258V}.

\begin{figure}
    \centering
    \includegraphics[width=0.8\linewidth]{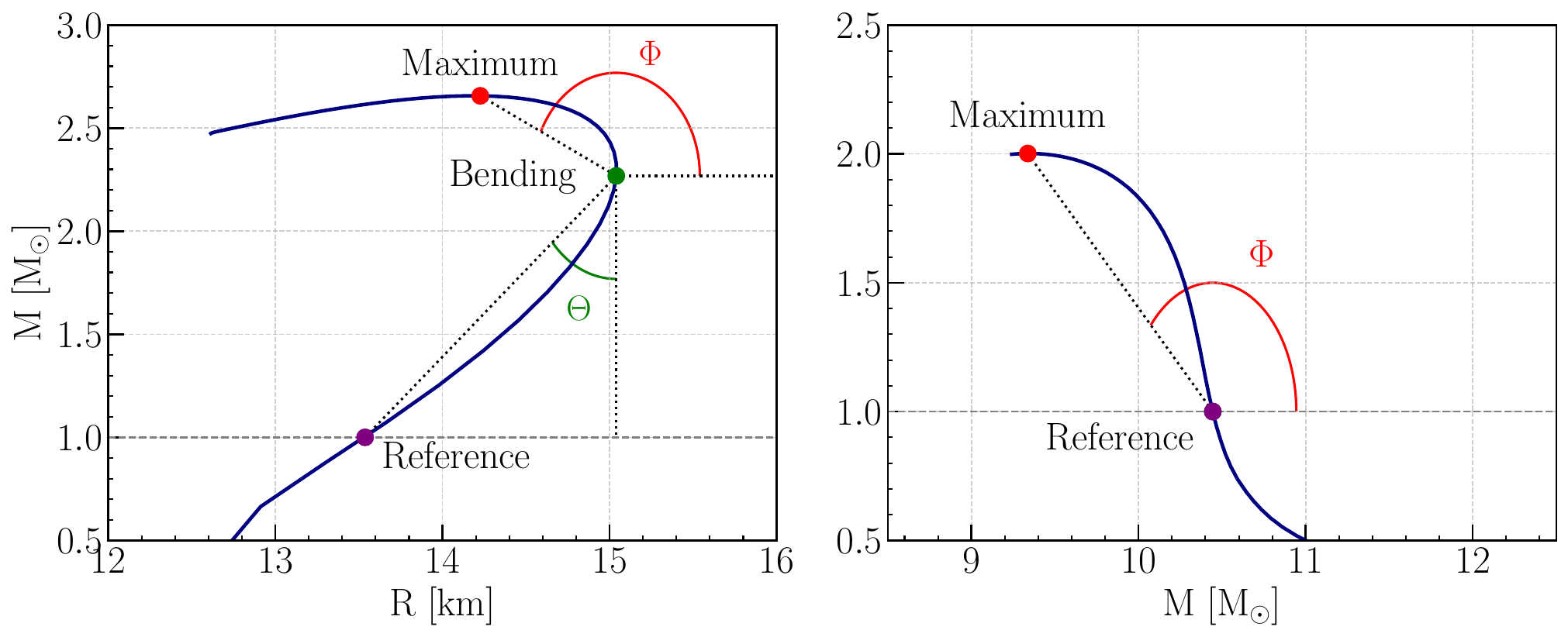}
    \caption{The proposed bending parameters, FB \& BB are defined based on the three points, namely ($\mathrm{M_{max},\,R_{max}}$), ($\mathrm{M_{bend},\,R_{bend}}$), and ($\mathrm{M_{ref},\,R_{ref}}$) that are found from the geometry of the M-R sequence. The front bending and back bending are defined as FB = $\cot \Phi$ and BB = $\cot \Theta$. In the figure, a typical M-R sequence with back-bending and front-bending is depicted, along with their characterisation in the left and right panels, respectively.}
    \label{BB-FB}
\end{figure}

Once the EoS ensemble is constructed, the TOV equation is solved to obtain an ensemble of M-R sequences. We intend to solve both the \textit{Maximum mass problem and the Inverse problem} using the geometric parameters. To achieve the goal, three characteristic points and two geometric parameters (FB and BB) are defined to characterise the M-R sequence. The three points are the maximum mass point ($\mathrm{M_{max},\,R_{max}}$), the bending point ($\mathrm{M_{bend},\,R_{bend}}$) (point on the M-R curve where $\mathrm{\frac{dR}{dM}\,=\,0}$) and the reference point ($\mathrm{M_{ref},\,R_{ref}}$) of the M-R sequence. 
The presence of a bending point in an M-R curve signifies the phenomenon of back-bending, while its absence suggests front-bending. The reference point is a fixed point at $\mathrm{M_{ref}\,=\,1\,M_{\odot}}$, and the $\mathrm{R_{ref}}$ is the radius of the star at that mass $\mathrm{M_{ref}}$ on the M-R curve. In case of an M-R where the bending point is absent, the bending and the reference point are taken to be the same, producing only the front-bending angle. The two parameters, FB and BB, quantify the degree of bending in an M-R sequence. 

We define the FB = $\cot \Phi$, and the BB = $\cot \Theta$ as shown in Fig. \ref{BB-FB}. These two parameters theoretically have bounds, $\mathrm{FB \in\, (-\infty, \,0]}$, and $\mathrm{BB \in\, [0, \,\infty)}$. Among these two parameters, the BB parameter, having higher Mutual Information (MI), contains more information about the M-R sequence than the FB parameter. Table~\ref{MI} in Appendix C summarises the MI values between the maximum mass point and different combinations of bending and reference parameters, evaluated separately for monotonic and non-monotonic $c_s^2$ profiles. 

\begin{figure}[H]
    \centering
    \includegraphics[width=0.8\linewidth]{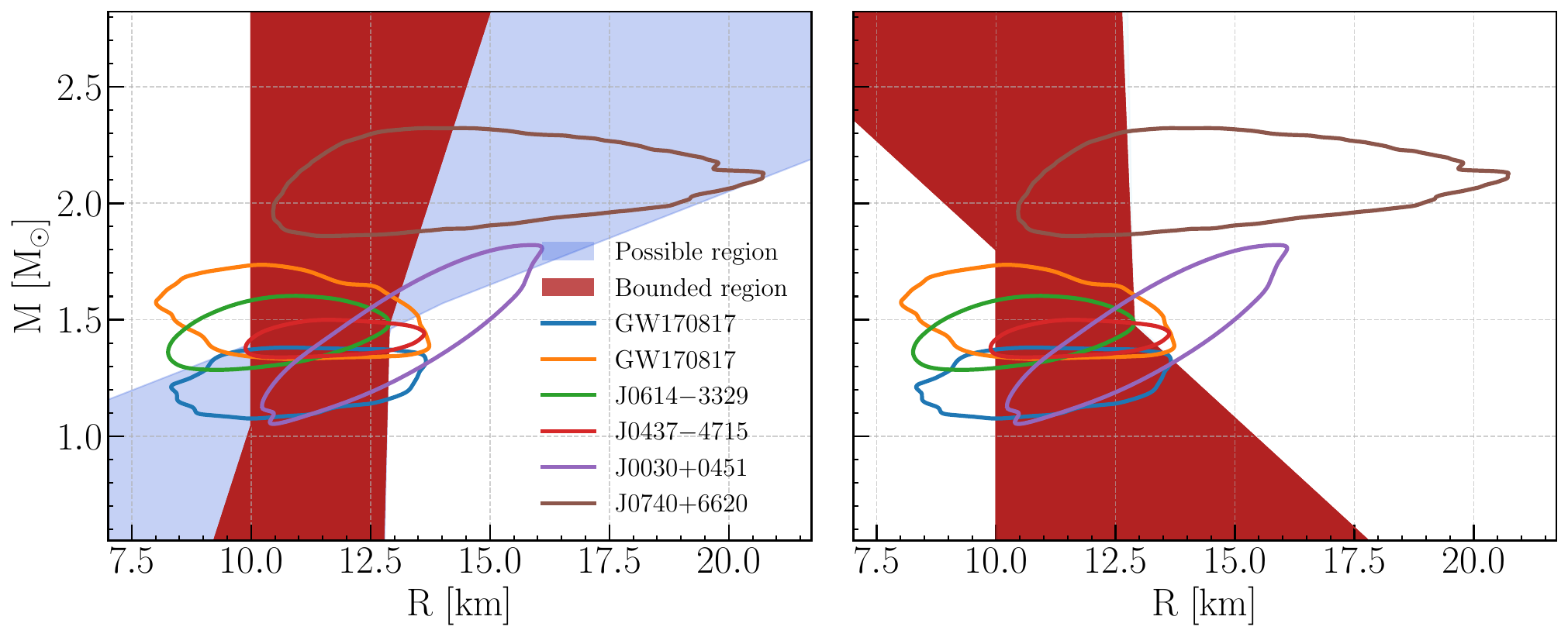}
    \caption{Connecting astrophysical observations with the bending parameters. (Left panel) The probable region of M-Rs with BB that satisfy observations is depicted. (Right panel) The probable region of M-Rs with FB that satisfies observations is shown. The bounded regions are obtained by imposing bounds on the bending parameters based on observations.}
    \label{valid_region}
\end{figure}

The M-R sequences are of two distinct classes, one corresponding to $BB = 0$, representing FR M-R sequences whose geometry is governed by FB and the reference point, and the other to $BB \neq 0$, representing BB M-R sequences whose geometry is governed by BB and the reference point. For these two classes, we select the parameter sets $[BB, M_{\mathrm{ref}}, R_{\mathrm{ref}}]_{BB \neq 0}$ and $[FB, M_{\mathrm{ref}}, R_{\mathrm{ref}}]_{BB = 0}$, respectively, to infer the probable maximum mass of the NS. Since these inputs depend solely on the geometry of the allowed region in the M-R plane, we incorporate the present observational constraints to determine the admissible regions, as shown in \cref{valid_region} (the observational contours correspond to a $95\%$ confidence level). The bounds on the bending parameters from the agnostic data  $BB \in [0,\,1.617]$ and $FB \in [-0.188,\,-5.016]$ are used to restrict the possible region in the M-R space. To increase the efficiency of scanning the M-R space, the allowed region is relaxed, with the lower boundary set to 10 km.

\begin{figure}[H]
    \centering
    \includegraphics[width=1.02\linewidth]{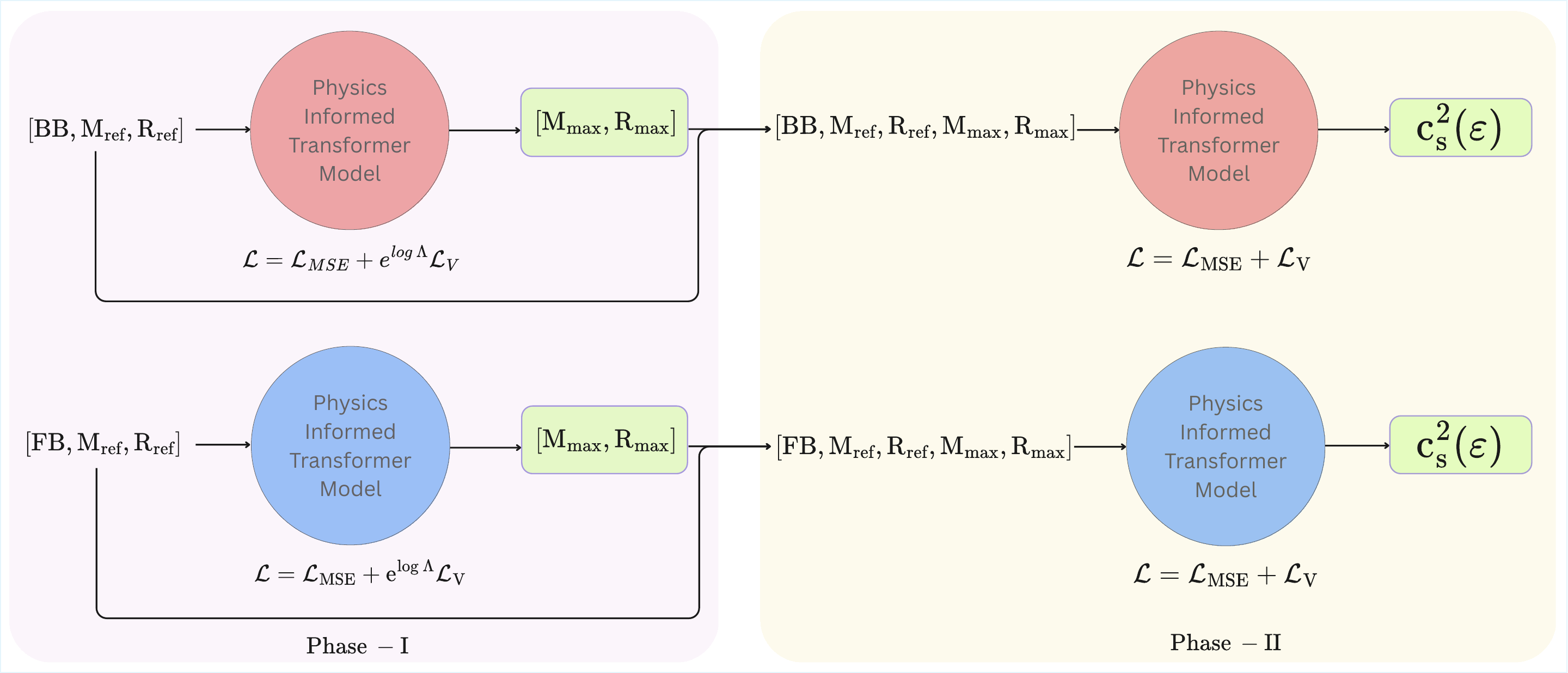}
    \caption{A schematic representation of the unified transformer model framework showing the two-step network: First, the forward solution to obtain $\mathrm{(M_{max},R_{max})}$ and then the inverse reconstruction of $c_s^2$ from NS observables.}
    \label{flow}
\end{figure}

The correlation between the NS observables and the $c_{s}^2$ is highly non-linear in nature, and restricts one to propose any kind of quasi-empirical relation. 
So we move to a Physics-Informed Machine Learning approach. All four units (as shown in \Cref{flow}) are based on Feature Tokeniser Transformer (FT-Transformer) architecture, specifically designed for tabular data \cite{DBLP:journals/corr/abs-2106-11959}. Physics-informed loss functions (virtual losses) are added to the Transformer models \cite{2025NatRP...7..154A} in each of the four steps (as shown in Fig. \ref{flow}). The details of the Transformer model architecture, working principle, and loss functions are discussed in Appendix \cref{Arc}.

Our agnostically M-R sequences are of two classes: BB and FB. For each class, we first address the maximum mass from the governing variables of that class, and then, in the next step, predict the $c_s^2(\varepsilon)$. The training PINN model is structured into two distinct phases, as illustrated in \cref{flow}. In the first phase, we train our models to predict the maximum mass point $[M_{\mathrm{max}}, R_{\mathrm{max}}]$ directly from the geometric inputs derived from the allowed region in the M-R space. In the second phase, the inferred maximum mass points are combined with the initial geometric inputs to form an augmented feature set for predicting the square of the NS speed-of-sound profile. The second phase essentially solves the \textit{inverse problem}, mapping from M-R sequence geometry back to the underlying equation-of-state (EoS) information. 

\section{Results}\label{sec2}

\begin{figure}[H]
    \centering
    \includegraphics[width=0.8\linewidth]{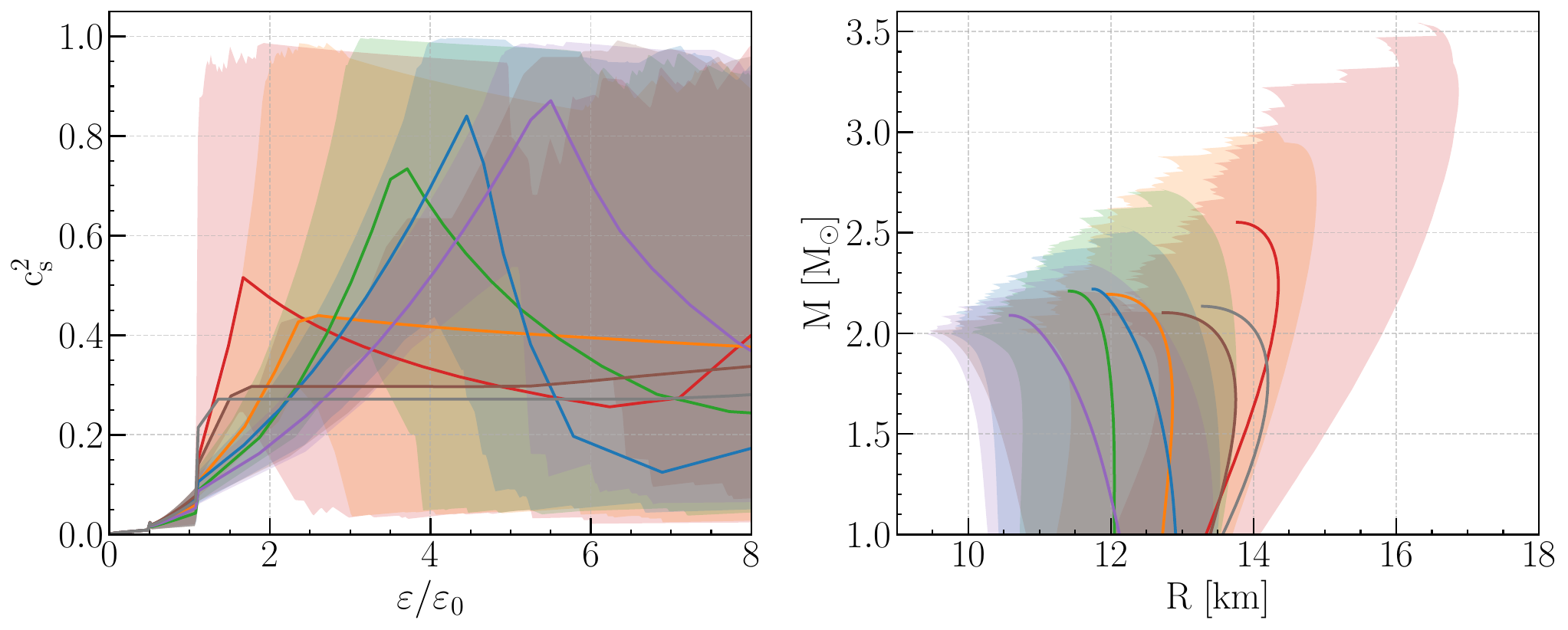}
    \caption{The figure shows the $c_s^2$ peak profile as energy density and the M-R sequence coded as $c_s^2$ peak profile. The rise in $c_s^2$ after the CET band dictates the bending properties in the M-R sequence. (Left panel) The $c_{s}^{2}$ vs $\varepsilon$ plot where the energy density is normalised with respect to the nuclear saturation energy density. This figure shows the distribution of the $c_s^2$ peak values in the agnostic data. (Right panel) M-R sequences of the corresponding groups of the square of the speed of sound profiles. The ${c_s}^2$ profiles have been grouped by the location of their maxima, and this shows that bending in the M-R sequence depends on the location of the first maxima of ${c_s}^2$.}
    \label{eos-mr}
\end{figure}

Matter properties are better illustrated by the speed of sound, which is the derivative of the EoS. As one tries to map the M-R sequence to the EoS, the $c_s^2$ captures this efficiently, as shown in Fig. \ref{eos-mr}. Different-coloured regions in the left panel indicate the position of the peak of $c_s^2$ across various energy-density ranges. The right panel shows the corresponding M-R sequence contours. The EoS whose $c_s^2$ peaks at low density produces more massive stars than those whose peak appears at higher density. The bending parameters FB, BB are more sensitive to the location of the maximum of the $ c_s^2 (\varepsilon)$ than its magnitude, where, on the other hand, the maximum mass is affected by both the bending parameters as well as the maximum value of $c_s^2$. Thus, the characteristics of the M-R sequence are actually produced by a beautiful interplay of the features of $c_s^2(\varepsilon)$. The EoS with a relatively low density $c_s^2(\varepsilon)$ peak shows significant BB, and the EoS with a significantly high density peak shows significant FB. For the EoS, which peaks at intermediate densities, the distinction is not very clear. As a result, curves with significant BB can generate massive stars, whereas M-R sequences showing high FB can yield only masses that can just satisfy the present lower maximum mass cut-off.

\begin{figure}[H]
    \centering
    \includegraphics[width=0.8\linewidth]{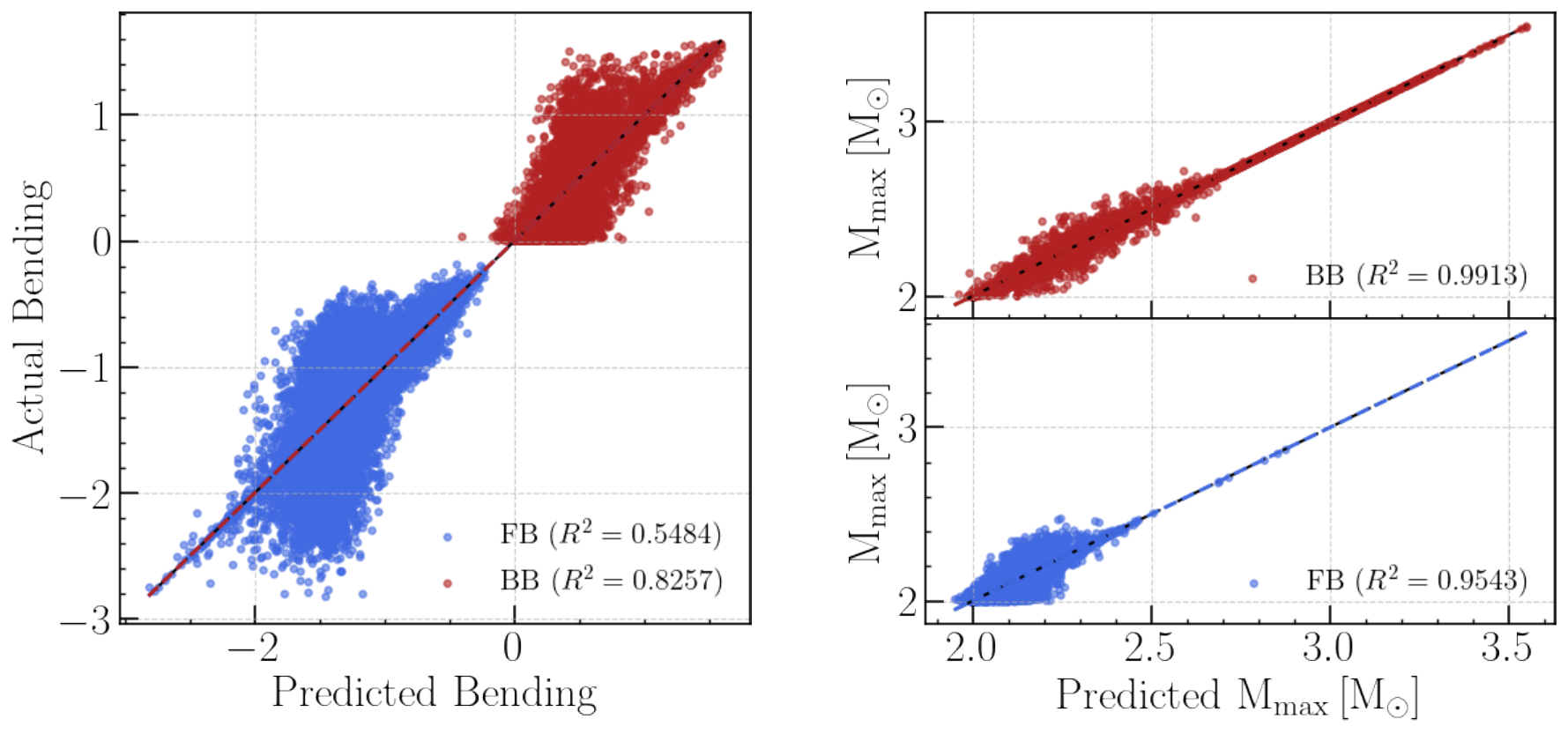}
    \caption{The figure shows the distribution of the bending parameters and the amount of information carried by them about the maximum mass of NSs. (Left panel) The distribution of the actual bending parameters vs the predicted values, where we used a Fourier fit of $n_{F}\,=\, 120$. (Right panel) The actual vs predicted values of $\mathrm{M_{max}}$, where we used $n_{F}\,=\,120$ }
    \label{fourier-fit}
\end{figure}

The bending parameters depend on the features of the $c_s^2$ profile, i.e., the maximum of $c_s^2=c_{s, max}^2$, and the value of energy density at which the $c_s^2$ has the maxima ($\varepsilon_{\mathrm{peak}}$). The $\mathrm{M_{max}}$ depends on the bending parameters, as well as on $c_{2, max}^2$, $\varepsilon_{\mathrm{peak}}$. The empirical expressions are given by, 

\begin{equation}
    \mathcal{Q} = \mathcal{Q}(c_{s,\max}^2,\varepsilon_{\mathrm{peak}})
    \label{fouriereq1}
\end{equation}

\begin{equation}
    \hat{Y}_{\mathcal{Q}} = \hat{Y}(\mathcal{Q}, c_{s,\max}^2,\varepsilon_{\mathrm{peak}})
    \label{fouriereq2}
\end{equation}

where $\mathcal{Q} \in [BB, FB]$, and $\hat{Y}_{\mathcal{Q}}$ is the $\mathrm{M_{max}}$.

Fig. \ref{fourier-fit} (left panel) depicts the distribution of the bending parameters. The distribution is non-localised; however, the BB shows a relatively compact distribution with a high $R^2$ score. As a result, the dependence of the maximum mass on the BB score is higher than on the FB score, as depicted in the right panel of the figure, because BB carries more MI about the M-R sequence. The detailed expression of these empirical forms of the Fourier fit is discussed later in the Appendix \ref{appfourier}. 

\begin{figure}[H]
    \centering
    \includegraphics[width=0.8\linewidth]{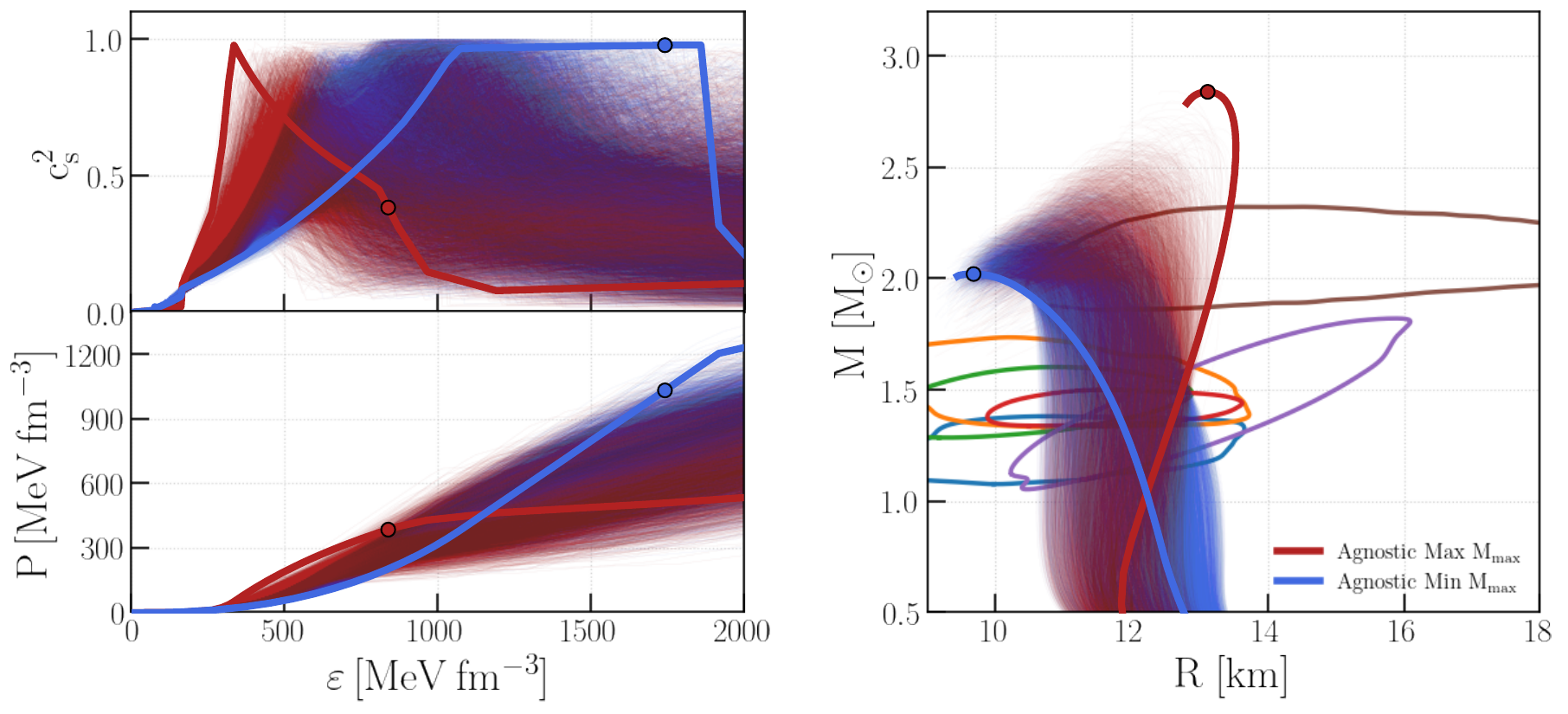}
    \caption{(Left panel)The plots show the EoS and $c_s^2$ for the agnostic ensemble. (Right panel) The M-R sequences from the agnostic data which lie inside the observation contours. The bold red and blue curves show the extremal values of the M-R sequence for the entire ensemble, along with their corresponding EoS and ${c_s}^2$.}
    \label{agnostic}
\end{figure}

Before we discuss the results of the maximum mass and the \textit{Inverse problem} from the Transformer model, let's analyse the EoS, $c_s^2$ profile and M-R sequence from the agnostic data. This would provide a better basis for comparison. 
Fig. \ref{agnostic} shows their EoS, $c_s^2$ profile, and the M-R sequences, classified in terms of the FB and BB scores, satisfying all the present astrophysical bounds. 
As expected, the BB EoS has its $c_s^2$ peak at lower densities, and the EoS exhibits high stiffness at low densities and softens at high densities. On the other hand, the FB $c_s^2$ peak occurs at higher density, so the EoS stiffens only at higher densities.

The right panel of Fig. \ref{agnostic} for the M-R sequence shows that comparatively low maximum masses are dominated by FB M-R sequences, whereas high maximum masses are dominated by BB M-R sequences. The maximum mass and its corresponding M-R sequence of the entire set is shown by the bold red curve. It can generate an NS as massive as $2.845$ solar masses with a radius of $13.116$ km. The corresponding EoS is very stiff at low densities and softens drastically at high densities. The $c_s^2$ peaks at very low density and then falls off sharply. Thus, the universal fact is that to generate massive stars, the $c_s^2$ must peak at low density, thereby making the EoS very stiff at low density. Interestingly, the EoS producing the highest maximum mass also has the highest BB score. The EoS with the minimum FB has the minimum possible mass, but it is not the EoS with the minimum radius. The minimum radius of NS for a $1.4$ solar mass star is approximately $10.627$ km, and for a $2$ solar mass star, it is $9.314$ km. Looking at the EoS plot, one observes that they must be very soft at low density, whereas their high-density behaviour matters little.

\begin{figure}[H]
    \centering
    \includegraphics[width=0.8\linewidth]{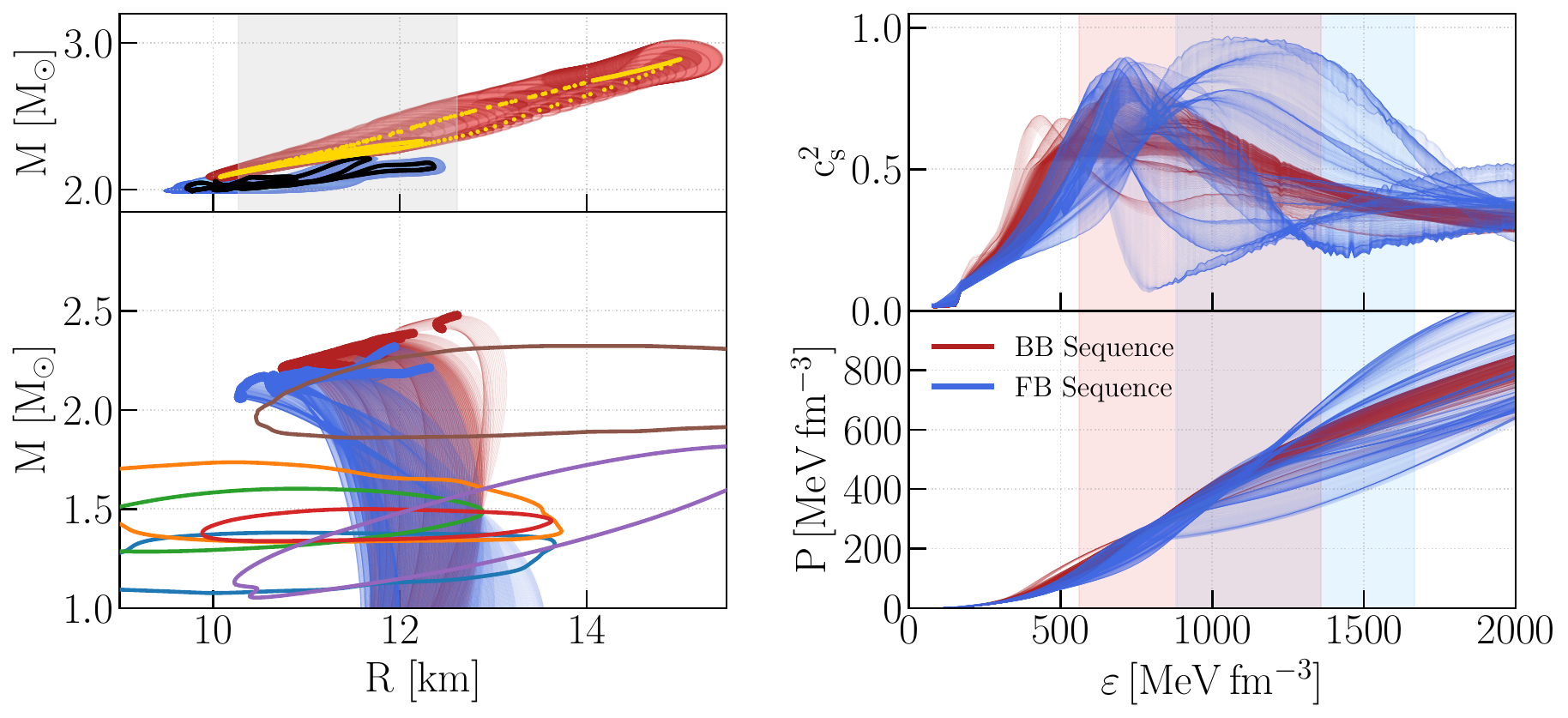}
    \caption{(Top left panel) The predicted $\mathrm{(M_{max}, R_{max})}$ from the step-1 of the Transformer model is shown in the plot. The yellow points are the predictions from the BB class, and the black dots are the predictions from the FB class. The deep red contours are the 1 $\sigma$ and the pale red contours are the 2 $\sigma$ zones of the BB predictions. Similarly, the blue region in the FB class prediction represents the uncertainty, where deep blue corresponds to the 1 $\sigma$ and light blue to the 2 $\sigma$ uncertainty contours. The grey-shaded region represents the extremum in the $\mathrm{(M_{max}, R_{max})}$ data that satisfies the observational constraints. (Top right panel) The \textit{inverse problem} predicted of the $c_s^2(\varepsilon)$ profile from our physics-informed Transformer model. The red and the blue curves represent the BB and FB classes, respectively. (Bottom right panel) The reconstructed EoS from the predicted $c_s^2(\varepsilon)$ profile. (Bottom left panel) The reconstructed M-R sequences satisfying all the observational constraints, from the reconstructed EoS.}
    \label{transformer}
\end{figure}

Fig. \ref{transformer} [left panel] shows the maximum mass (and radius) prediction by the Transformer model. The maximum mass predicted by the transformer model for the sequences is around $2.889$ solar masses, similar to the agnostic maximum mass for NSs. However, when observational constraints are applied, it is reduced to $2.477$ solar masses, a value similar to those reported earlier \cite{Rezzolla_2018}. Additionally, the maximum mass sequence does not have the maximum BB score, a deviation from agnostic data. The fact that the BB score carries more information is also evident from the figure. Almost the entire maximum mass region is covered by the BB sequences, while the FB sequences are distributed only at low maximum mass values.

The results of the solution of the \textit{inverse problem} are shown in the right panel of Fig. \ref{transformer}. First, the $c_s^2$ sequences are obtained, and then the EoS is determined. The BB sequences show a peak at low density and generate massive stars, whereas the FB sequences show a peak at high density and generate less massive stars. The EoS that produces the maximum mass is stiffer at low densities and substantially softer at high densities. The sequences with the maximum FB that yield the lowest maximum mass are softer at low density but substantially stiffer at high density. The density at which it attains the maximum mass is also substantially high. 

Although the overall feature remains more or less the same, where BB EoS generates more massive stars, and FB generates less massive stars, several features deviate from agnostic M-R sequences. The curve with the maximum FB, which generates a 2 solar mass star, has a radius of approximately $10.273$ km. The minimum radius for a canonical NS of $1.4$ solar mass star is around $11.5$ km. None of the EoS (from both the agnostic and ML models) can reach the minimum radius set by astrophysical constraints. This is because of the interplay of the minimum radius set by PSR J0740+6620 at higher mass and the maximum value of FB dictated by causality.

\section{Summary and Discussion}

In this article, we addressed the maximum mass of NSs and the \textit{inverse problem} of NSs. As the EoS of high-density matter is beyond the reach of theory or experiment, we employ the inverse method, in which one uses NS observations to map the EoS. We employ a transformer-based ML model to perform the inverse mapping. The mapping is primarily guided by astrophysical observations of NSs through NICER and GW170817. To address this problem, we employ a large ensemble of EoS that fully exploit the theoretical freedom available to us, thereby minimising biases. 

Once the ensemble of EoS is available, it is fed into the TOV equation to obtain an ensemble of M-R sequences. To relate the M-R sequence to the properties of matter, specifically the square of the sound speed and energy density, we define two characteristics of the M-R curve, the FB and the BB. Once the characterisation is done, we train our model. The model delivers a two-step result. First, the model predicts the maximum mass and its corresponding radius. In the next step, along with the initial input, the output of the first step is used to predict the square of the sound speed as a function of density, completing the inverse mapping.

The BB is seen to carry more information and therefore gives more accurate results. The FB sequences need more guidance from the physics-informed constraints. The maximum mass predicted by the transformer model without any astrophysical constraints is around $2.9$ solar masses, but drops to $2.477$ solar masses once the constraints are applied. The \textit{inverse problem} predicts that the most massive stars are produced by EoS sequences with higher BB scores, where the $c_s^2$ peaks at low density. FB sequences dominate the less massive M-R sequences, where the $c_s^2$ peaks at relatively higher densities. The EoS for which the M-R sequences generate very massive stars are very stiff at low densities, and the core central density is comparatively small, unlike in low-mass stars.

Traditionally, machine-learning approaches in neutron-star physics proceed in a forward manner, wherein an EoS is first used to construct M-R sequences, a model is trained on these forward mappings, and the EoS is subsequently inferred from a given M-R curve. However, for a fixed set of observational constraints, infinitely many M-R curves can be constructed that satisfy all observations, rendering the inverse inference highly degenerate in the forward-learning paradigm. In contrast, our formalism adopts a fundamentally different approach by performing an inverse mapping that depends solely on the M-R sequence's geometry. Given the observational constraints, the allowed region in the M-R plane is bounded, thereby restricting the possible reference points and bending parameters. This geometric restriction imposes a finite and well-defined domain over which the corresponding EoS can be traced. As additional NS M-R observations become available, the allowed region in the M-R space would shrink further, progressively narrowing the range of viable EoS models within our framework.

According to our analysis, the maximum mass allowed by present astrophysical constraints for an NS is around $2.889$ solar masses. Recently, a few objects lying in the mass gap region have been observed with the LIGO-VIRGO-KAGRA collaboration \cite{Kacanja_2025}. According to our analysis, any object with a static mass greater than $2.8$ solar masses is definitely a BH; an object with a mass between $2.477$ and $2.889$ solar masses is ambiguous; and an object with a mass less than $2.477$ solar masses is definitely a NS. The number quoted here is an outcome of the present astrophysical bounds from NICER and GW170817. In the future, more and better observations can either confirm this maximum mass or push it down further. However, the upper bound on the mass is unlikely to increase further.

\section{Acknowledgments}

The authors RM and RM thank the Indian Institute of Science Education and Research Bhopal for providing all the research and
infrastructure facilities, and the HPC BHASKARA. RM acknowledges the Science and Engineering Research Board (SERB), Govt. of India, for financial support in the form of a Core Research Grant (CRG/2022/000663). 

\section{Data Availability}

The data for the work are available at Zenodo and figshare.

\bibliography{Papers}

\appendix

\section{Architecture of the Physics Informed Transformer Model}
\label{Arc}

\begin{figure}[H]
    \centering
    \includegraphics[width=1\linewidth]{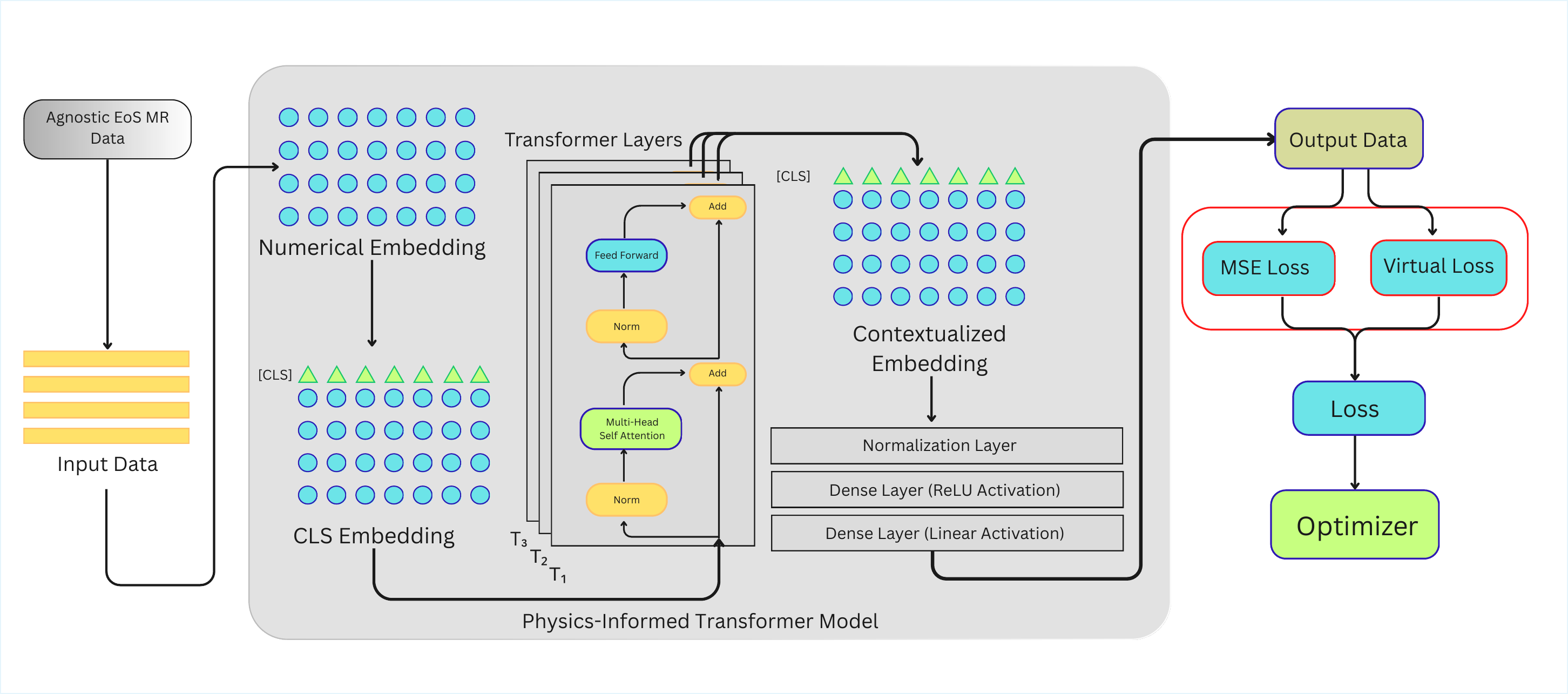}
    \caption{A schematic diagram showing the details of the architecture network of the physics-informed Transformer model}
    \label{fig:placeholder}
\end{figure}

The primary characteristic of the Feature Tokeniser Transformer (FT Transformer) model is the Feature Tokenisation layer. Unlike standard neural networks that operate on raw numerical vectors, the FT-Transformer projects each individual feature $x_i$ into a high-dimensional latent space $\mathbb{R}^{n}$. This transformation is defined as:

\begin{equation}
    \mathbf{T}_i = x_i \cdot \mathbf{W}_i + \mathbf{b}_i
\end{equation}

where $\mathbf{W}_i \in \mathbb{R}^{1 \times n}$ and $\mathbf{b}_i \in \mathbb{R}^{n}$ are learnable weights and biases specific to each feature. By embedding each physical parameter into its own n-dimensional vector, the model can capture each variable's unique "context" before modelling their interactions. At the beginning of the sequence, we prepend a learnable d-dimensional vector $\mathbf{z}_{CLS} \in \mathbb{R}^{n}$ that does not represent any physical input. As the sequence passes through a stack of $L$ number of Transformer layers (optimised via hyperparameter tuning), the Classification ([CLS]) token interacts with the physical tokens via the Multi-Head Self-Attention (MHSA) mechanism. Through these layers, the [CLS] token "attends" to the entire feature set, effectively acting as a global aggregator of information. It learns to weigh the importance of different feature combinations. After the final Transformer block, the state of the [CLS] token is extracted and passed through a specialised output head comprising a Normalisation Layer and two Dense layers (employing ReLU and Linear activations, respectively). This ensures the final prediction of output is a representation derived from the collective feature context.

In Phase I, while predicting the $\mathrm{(M_{max}, R_{max})}$, to incorporate observational bounds of the allowed bounded regions in the M-R space, \textit{Physics-Informed Loss Function} is implemented. The total Loss-function $\mathcal{L}$ is a sum of the empirical error (Mean Squared Error Loss) and a physical penalty term (Virtual Loss):

\begin{equation}
    \mathcal{L_{I}} = \mathcal{L}_{MSE} + e^{\log \Lambda} \cdot \mathcal{L}_{V}
\end{equation}

The term $\mathcal{L}_{V}$ (Virtual Loss) represents a set of physical boundary constraints in the M-R plane. The allowed regions, as shown in \cref{valid_region}, have been used to train the model. Penalties were imposed if the $\mathrm{(M_{max}, R_{max})}$ was outside the allowed region. The weighting factor $\Lambda$ was treated as a learnable parameter, and we used the form $e^{\log \Lambda}$ so that the algorithm doesn't make the $\Lambda\,<\,0$ in order to minimise the total loss. 

In Phase II, while predicting the $c_{s}^{2}$, another set of physics constraints were imposed on the $\mathcal{L}_{V}$; Strong penalties were imposed if the maximum value of the $c_{s}^2(\varepsilon)$ exceeded the causality limit or was less than 0.33 (conformal limit) (for both the BB, and the FB models). For the FB model, since the FB parameter contains less information, two additional checks were imposed on the loss function. From the agnostic data, it was found that for an FB M-R, $c_{s}^2(\varepsilon)$ should be strictly monotonically increasing in the lower energy-density range [\cref{eos-mr}]. Upon plotting all the FB M–Rs and their corresponding $c_{s}^{2}(\varepsilon)$ profiles, it was observed that the $c_s^2$ remained monotonically increasing up to $500,\mathrm{MeV,fm^{-3}}$. Based on this systematic behaviour, penalties were imposed during the training process whenever $c_{s}^{2}(\varepsilon)$ exhibited non-monotonic behaviour within $500,\mathrm{MeV,fm^{-3}}$. Additional penalties were also imposed when $c_{s}^{2}(\varepsilon = 170,\mathrm{MeV,fm^{-3}})$ exceeded $0.1161$, and if $c_{s}^{2}(\varepsilon = 5000,\mathrm{MeV,fm^{-3}})$ exceeded $0.7014$. For Phase II, we had set the value of $\Lambda\, = \, 1$, and the total loss was given by,

\begin{equation}
    \mathcal{L_{II}} = \mathcal{L}_{MSE} + \mathcal{L}_{V}
\end{equation}

To statistically assess the model's prediction uncertainty, \textit{Monte Carlo Dropout} was employed for each Phase I prediction, specifically for predicting the maximum mass–radius points. During the inference phase, the Dropout layers, with rates between 0.1 and 0.4, were kept active. For each unique input, 100 stochastic forward passes were performed to generate a predictive distribution. The mean of this distribution was taken as the predicted $\mathrm{(M_{max}, R_{max})}$, while the standard deviation ($\sigma$) was interpreted as the associated epistemic uncertainty ($\sigma_{E}$). To additionally account for the aleatoric uncertainty ($\sigma_{A}$), the entire procedure was repeated 1,000 times for the same input features, and the final mean over all repetitions was defined as the final prediction. The total uncertainty ($\sigma^{i}$) for both $\mathrm{M_{max}}$ and $\mathrm{R_{max}}$ was then obtained by combining the individual uncertainty components using the following relation:

\begin{equation}
    \sigma^{i} = \sqrt{\sigma_{E}^{2} + \sum_{j=1}^{1000} \sigma_{A_{j}}^{2}}
    \label{stat}
\end{equation}

In this formulation, $i$ denotes the specific physical parameter ($\mathrm{M_{\text{max}}}$ or $\mathrm{R_{\text{max}}}$), and $j$ represents each individual uncertainty recorded across the 1,000 separate prediction runs for that same input.

For Phase II models, we used an Ensemble learning method: we trained 5 simultaneous models to estimate epistemic and aleatoric uncertainties. To calculate the total uncertainty, we used the same equation given in \cref{stat}.

All four PINN units (as shown in \cref{flow}) were optimised using the AdamW algorithm. Hyperparameter optimisation was conducted using the Optuna framework, which employed a 5-fold cross-validation scheme to determine the optimal values for the number of blocks, learning rate, and dropout probability. To ensure convergence, we employed a ReduceLROnPlateau scheduler and an Early Stopping mechanism.

\section{Coupled Fourier Equations}
\label{appfourier}

In order to establish a relation between the Bending parameters with the $c_s^2$ features, we used a coupled Fourier fit \cref{fouriereq1}. The explicit function is given by,

\begin{equation}
\label{eq:fourier_inverse_master}
\begin{aligned}
\mathcal{Q}(c_{s,\max}^2,\varepsilon_{\mathrm{peak}})
\;=\;& a_0
\\
&+ \sum_{k=1}^{n}
\left[
a_k \cos\!\left(k\,c_{s,\max}^2\right)
+ b_k \sin\!\left(k\,c_{s,\max}^2\right)
\right]
\\
&+ \sum_{\ell=1}^{n}
\left[
c_\ell \cos\!\left(\ell\,\varepsilon_{\mathrm{peak}}\right)
+ d_\ell \sin\!\left(\ell\,\varepsilon_{\mathrm{peak}}\right)
\right]
\\
&+ \sum_{k=1}^{n}\sum_{\ell=1}^{n}
\Big[
e_{k\ell}
\cos\!\left(k\,c_{s,\max}^2\right)
\cos\!\left(\ell\,\varepsilon_{\mathrm{peak}}\right)
\\
&\hspace{2.6cm}
+ f_{k\ell}
\cos\!\left(k\,c_{s,\max}^2\right)
\sin\!\left(\ell\,\varepsilon_{\mathrm{peak}}\right)
\\
&\hspace{2.6cm}
+ g_{k\ell}
\sin\!\left(k\,c_{s,\max}^2\right)
\cos\!\left(\ell\,\varepsilon_{\mathrm{peak}}\right)
\\
&\hspace{2.6cm}
+ h_{k\ell}
\sin\!\left(k\,c_{s,\max}^2\right)
\sin\!\left(\ell\,\varepsilon_{\mathrm{peak}}\right)
\Big]
\end{aligned}
\end{equation}

Where all the parameters $a_o, a_k, b_k, c_l, d_l, e_{kl}, f_{kl}, g_{kl}, h_{kl}$ have been estimated, for the Fourier fit of $n_F$ = 120. Similarly, we used the same approach for finding the fit for the relation between the maximum mass and the bending parameters \& $c_s^2$ features using the coupled Fourier fit, where the empirical relation \cref{fouriereq2} is given by,

\begin{equation}
\setlength{\jot}{0pt}
\begin{aligned}
\hat{Y}(x,y,z) =\;& a_0 
+ \sum_{k=1}^{N_F} \Big[
a_k^{(x)} \cos(kx) + b_k^{(x)} \sin(kx)
+ a_k^{(y)} \cos(ky) + b_k^{(y)} \sin(ky)
+ a_k^{(z)} \cos(kz) + b_k^{(z)} \sin(kz)
\Big] \\
&+ \sum_{k,m=1}^{N_F}
\Big[
c_{km}^{(xy,cc)} \cos(kx)\cos(my)
+ c_{km}^{(xy,cs)} \cos(kx)\sin(my)
+ c_{km}^{(xy,sc)} \sin(kx)\cos(my)
+ c_{km}^{(xy,ss)} \sin(kx)\sin(my)
\Big] \\
&+ \sum_{k,m=1}^{N_F}
\Big[
c_{km}^{(xz,cc)} \cos(kx)\cos(mz)
+ c_{km}^{(xz,cs)} \cos(kx)\sin(mz)
+ c_{km}^{(xz,sc)} \sin(kx)\cos(mz)
+ c_{km}^{(xz,ss)} \sin(kx)\sin(mz)
\Big] \\
&+ \sum_{k,m=1}^{N_F}
\Big[
c_{km}^{(yz,cc)} \cos(ky)\cos(mz)
+ c_{km}^{(yz,cs)} \cos(ky)\sin(mz)
+ c_{km}^{(yz,sc)} \sin(ky)\cos(mz)
+ c_{km}^{(yz,ss)} \sin(ky)\sin(mz)
\Big] \\
&+ \sum_{k,m,l=1}^{N_F}
\sum_{f_1,f_2,f_3 \in \{\cos,\sin\}}
d_{kml}^{(f_1 f_2 f_3)} f_1(kx) f_2(my) f_3(lz).
\end{aligned}
\end{equation}

\vspace{2cm}

All the parameters were estimated for a Fourier fit of degree $n_F$ = 20.

In order to establish the accuracy of the fit, we used the measure $R^2$ or the coefficient of determination, which is defined by,

\begin{equation}
R^2 = 1 - \frac{\sum_{i=1}^{N} \left( y_i - \hat{y}_i \right)^2}
{\sum_{i=1}^{N} \left( y_i - \bar{y} \right)^2},
\end{equation}

where $y_i$ are the true values, $\hat{y}_i$ are the predicted values, and $\bar{y} = \frac{1}{N}\sum_{i=1}^{N} y_i$, $\bar{y}$ is the mean of the observed data.

\section{Tables}

\subsection*{Mutual Information Table}

\begin{table}[!htbp]
\renewcommand{\arraystretch}{1.15}
\begin{tabular}{@{}l l r@{}}
\toprule
\textbf{Observable Pair} & \textbf{$c_s^2$ Profile} & \textbf{MI} \\
\midrule
$MI(M_{\max}; \mathrm{FB})$ 
& Monotonic 
& 0.0266 \\
& Non-monotonic 
& 0.1146 \\
\addlinespace

$MI(M_{\max}; \mathrm{BB})$ 
& Monotonic 
& 0.1636 \\
& Non-monotonic 
& 0.3712 \\
\addlinespace

$MI((M_{\max}, R_{\max}); \mathrm{FB})$ 
& Monotonic 
& 0.2472 \\
& Non-monotonic 
& 0.3117 \\
\addlinespace

$MI((M_{\max}, R_{\max}); \mathrm{BB})$ 
& Monotonic 
& 0.6814 \\
& Non-monotonic 
& 1.1784 \\
\addlinespace

$MI((M_{\max}, R_{\max}); [\mathrm{BB}, M_{\mathrm{ref}}, R_{\mathrm{ref}}])$ 
& Monotonic 
& 1.0906 \\
& Non-monotonic 
& 1.6206 \\
\addlinespace

$MI((M_{\max}, R_{\max}); [\mathrm{FB}, M_{\mathrm{ref}}, R_{\mathrm{ref}}])$ 
& Monotonic 
& 0.6564 \\
& Non-monotonic 
& 0.7539 \\
\bottomrule
\end{tabular}

\caption{Table depicting the MI between NS maximum-mass observables and bending-related geometric parameters, evaluated separately for monotonic and non-monotonic square of the speed-of-sound profiles.}
\label{MI}
\end{table}

The MI between different sets of input variables and the maximum mass point is listed above, where the expression of MI for two random variables is given by,

\begin{equation}
    MI(X; Y) = \sum_{x \in \mathcal{X}} \sum_{y \in \mathcal{Y}} p(x, y) \log \frac{p(x, y)}{p(x)p(y)}
\end{equation}

where, $p(x,y) = P(X = x, Y = y)$ denotes the joint probability mass function (PMF) of $X$ and $Y$, $p(x) = P(X = x)$ and $p(y) = P(Y = y)$ are the marginal probability mass functions of $X$ and $Y$. Mutual Information represents the degree of correlation between two sets of data.

\subsection*{$\mathrm{(M_{\max}, R_{max})}$ Data Table}

\begin{table}[!htbp]
\label{tab:mr_comparison}
\renewcommand{\arraystretch}{1.15}
\begin{tabular}{@{}l l c c@{}}
\toprule
\textbf{Category} & \textbf{Description} & $\boldsymbol{\mathrm{M_{\max}\ [M_\odot]}}$ & $\mathrm{\boldsymbol{R(M_{\max})\ [\mathrm{km}]}}$ \\
\midrule
\multirow{2}{*}{PINN Prediction} 
& Upper limit & 2.889 & 15.002 \\
& Lower limit & 1.992 & 9.715 \\
\addlinespace

\multirow{2}{*}{Bounded M-R sequences} 
& Upper limit & 2.477 & 12.615 \\
& Lower limit & 2.059 & 10.273 \\
\addlinespace

\multirow{2}{*}{Agnostic data} 
& Upper limit & 2.845 & 13.116 \\
& Lower limit & 2.025 & 9.675 \\
\bottomrule
\end{tabular}
\caption{Comparison of the $\mathrm{M_{max}, R_{max}}$ from the Physics-Informer Transformer model, the M-Rs constructed from the inverse solution, and the Agnostic data}
\end{table}

The table summarises all the data on the maximum mass. Although the Transformer predictions give us a very high mass limit, the M-R sequences obtained from the reconstructed EoS obtained from the \textit{inverse mapping} do not satisfy all the observations. Therefore, the maximum mass of the NS satisfying all the observational bounds comes down to 2.477 $\mathrm{M_{\odot}}$.

\subsection*{Classification Table of GW Observations}

\begin{table}[!htbp]
\renewcommand{\arraystretch}{1.2}

\begin{tabular}{@{}l c c c c@{}}
\toprule

\textbf{Event} & $\mathbf{\mathcal{M}\ [M_\odot]}$ & $\mathbf{q}$ & $\mathbf{m_1\ [M_\odot]}$ & $\mathbf{m_2\ [M_\odot]}$ \\
\midrule
GW170817 & $1.19752^{+0.00013}_{-0.00019}$ & $1.264^{+0.215}_{-0.236}$ & \textcolor{RoyalBlue}{$1.549^{+0.131}_{-0.154}$} & \textcolor{RoyalBlue}{$1.225^{+0.090}_{-0.131}$} \\

\addlinespace
\addlinespace

GW190425 & $1.48632^{+0.00064}_{-0.00103}$ & $1.327^{+0.519}_{-0.299}$ & \textcolor{RoyalBlue}{$1.971^{+0.372}_{-0.241}$} & \textcolor{RoyalBlue}{$1.485^{+0.216}_{-0.198}$} \\

\addlinespace
\addlinespace

GW190814 & $6.44146^{+0.04976}_{-0.08109}$ & $11.710^{+3.032}_{-5.257}$ & \textcolor{FireBrick}{$28.657^{+4.392}_{-8.619}$} & \textcolor{RoyalBlue}{$2.447^{+0.205}_{-0.658}$} \\

\addlinespace
\addlinespace

GW200105 & $3.56297^{+0.04092}_{-0.02674}$ & $3.767^{+0.931}_{-2.339}$ & \textcolor{FireBrick}{$8.277^{+1.200}_{-3.407}$} & \textcolor{RoyalBlue}{$2.197^{+0.180}_{-1.213}$} \\

\addlinespace
\addlinespace

GW200115 & $2.57372^{+0.00901}_{-0.01144}$ & $2.744^{+3.175}_{-0.690}$ & \textcolor{FireBrick}{$5.019^{+2.726}_{-0.747}$} & \textcolor{RoyalBlue}{$1.829^{+0.521}_{-0.251}$} \\

\addlinespace
\addlinespace

GW230518 & $2.93215^{+0.01504}_{-0.02181}$ & $3.813^{+3.274}_{-2.220}$ & \textcolor{FireBrick}{$6.858^{+2.941}_{-2.615}$} & \textcolor{RoyalBlue}{$1.798^{+0.416}_{-0.865}$} \\

\addlinespace
\addlinespace

GW230529 & $2.02437^{+0.00282}_{-0.00211}$ & $1.604^{+1.919}_{-0.564}$ & \textcolor{FireBrick}{$2.961^{+1.575}_{-0.592}$} & \textcolor{RoyalBlue}{$1.846^{+0.559}_{-0.432}$} \\

\bottomrule

\end{tabular}
\caption{Binary component masses derived from the chirp mass $\mathcal{M}$ and mass ratio $q$.
Quoted uncertainties are asymmetric and correspond to the upper and lower bounds reported in the paper \cite{Kacanja_2025}.}
\end{table}

The list of recent GW observations \cite{2017GW170817, Abbott_2020_GW190425, Abbott_2020_GW190814, 2021ApJ915, Abac_2024,  Kacanja_2025} for small and medium range compact star mergers is given in the table. The classification of the compact objects in the respective binaries, either NS or BH, is done in accordance with the mass limit obtained from our ML prediction. The objects highlighted in blue are the NS, the red are BH, and the green are inconclusive.

\end{document}